\documentclass[sigconf]{acmart}

\usepackage{multirow}
\copyrightyear{2018} 
\acmYear{2018} 
\setcopyright{iw3c2w3}
\acmConference[WWW' 18 Compaion]{The 2018 Web Conference Companion}{April 23--27, 2018}{Lyons, France}
\acmPrice{}
\acmDOI{10.1145/3184558.3186928}
\acmISBN{978-1-4503-5640-4/18/04.}
\begin{document}
\title{GPSP: Graph Partition and Space Projection based Approach for Heterogeneous Network Embedding}
\author{Wenyu Du$^{\dag}$, Shuai Yu$^{\dag}$, Min Yang$^{\dag}$,  Qiang Qu$^{\dag 1}$, Jia Zhu$^{\ddag}$}
\affiliation{%
  \institution{$^\dag$ Shenzhen Institutes of Advanced Technology, Chinese Academy of Sciences}
  \institution{$^\ddag$ School of Computer Science, South China Normal University}
}
\renewcommand{\shortauthors}{Du et al.}

\begin{abstract}
\noindent In this paper, we propose \textit{GPSP}, a novel \textbf{G}raph \textbf{P}artition and \textbf{S}pace \textbf{P}rojection based approach, to learn the representation of a heterogeneous network that consists of multiple types of nodes and links.
Concretely, we first partition the heterogeneous network into homogeneous and bipartite subnetworks.
Then, the projective relations hidden in bipartite subnetworks are extracted by learning the projective embedding vectors. Finally, we concatenate the projective vectors from bipartite subnetworks with the ones learned from homogeneous subnetworks to form the final representation of the heterogeneous network.
Extensive experiments are conducted on a real-life dataset. The results demonstrate that \textit{GPSP} outperforms the state-of-the-art baselines in two key network mining tasks: node classification and clustering$^2$.
\end{abstract}


\maketitle
\footnotetext[1]{Q. Qu is the corresponding author (qiang@siat.ac.cn). The work was partially supported by the CAS Pioneer Hundred Talents Program.}
\footnotetext[2]{Data and codes are available at: https://github.com/Ange1o/GPSP.}
\section{Introduction}
Network embedding, or network representation learning, is the task of learning latent representation that captures the internal relations of rich and complex network-structured data. 
Inspired by the recent success of deep neural networks in computer vision and natural language processing, several recent studies~\cite{perozzi2014deepwalk,tang2015line,dong2017metapath} propose to employ deep neural networks to learn network embeddings. For example, DeepWalk~\citep{perozzi2014deepwalk} adopts Skip-gram~\citep{mikolov2013efficient} to randomly generate walking paths in a network; and LINE~\citep{tang2015line} tries to preserve two orders of proximity for nodes: first-order proximity (local) and second-order proximity (global).  

Most existing studies focus on learning the representation of a homogeneous network that consists of singular type of nodes and relationships (links). However, in practice, many networks are often heterogeneous~\cite{DBLP:conf/ssdbm/QuLYJ14,dong2017metapath}, i.e., involving multiple types of nodes and relationships. The methods designed for homogeneous networks hardly learn the representations of such networks because they cannot distinguish different types of objects and relationships contained in the networks. Therefore, the learned representations lack heterogeneity behind the structural information.    

To alleviate the aforementioned limitation, we propose a \textbf{G}raph \textbf{P}artition and \textbf{S}pace \textbf{P}rojection based approach (\emph{GPSP}) to learn the representation of a heterogeneous network. First, an edge-based graph partition method is used to partition the heterogeneous network into two types of atomic subnetworks: i) homogeneous networks that contain singular type of nodes and relationships; ii) bipartite networks that contain two types of vertices and one type of relationship. Second, we apply classic network embedding models~\cite{perozzi2014deepwalk,tang2015line} to learn the representations of homogeneous subnetworks. Third, for each bipartite subnetwork, the hidden projective relations are extracted by learning the projective embedding vectors for the related types of nodes. Finally, \textit{GPSP} concatenates the projective node vectors from bipartite subnetworks with the node vectors learned from homogeneous subnetworks to form the final representation of the heterogeneous network.

The main contribution of our approach is threefold: 
i) we formalize the problem of bipartite network representation learning; ii) edge-type based graph partition and space projection are used to learn the representations of different types of nodes in different latent spaces; and iii) the experimental results demonstrate the effectiveness of \textit{GPSP} in network mining tasks. 

\section{Our Model}
The definitions of homogeneous network~\cite{perozzi2014deepwalk} and heterogeneous network~\cite{dong2017metapath} are adopted. A bipartite network is defined:  

\begin{definition}{\textbf{A Bipartite Network}}
is defined as a graph $G=(V,E)$ where $V=V_1\cup V_2$ and $E=E_{V_1V_2}$. $V_1$ and $V_2$ are two types of vertex sets. In bipartite network each edge $e_{v_1v_2}\in E_{V_1V_2}$ connects two different types of nodes $v_1\in V_1$ and $v_2\in V_2$.
\end{definition}

\paragraph{Edge-type based graph partition}

For a heterogeneous network $G$, we first build a type-table to record all types of relationships in the network. The network is then partitioned into a minimum number of subnetworks, where each subnetwork is either a homogeneous network or a bipartite network. 

\paragraph{Homogeneous network embedding}
For homogeneous subnetworks, we employ conventional embedding algorithms such as LINE and DeepWalk to learn \emph{homogeneous embeddings}. The \emph{GPSP} framework  with LINE and DeepWalk algorithms are recorded as GPSP-LINE and GPSP-DeepWalk, respectively.  

\paragraph{Bipartite network embedding}
Unlike homogeneous networks, each edge in bipartite networks connects two different types of nodes.
After learning the representations of objects $O$ and $P$ in two different homogeneous networks (in two different low-dimensional spaces), we could treat the relationship between objects $O$ and $P$ in the bipartite networks as the implicit projection between two low-dimensional spaces. Based upon the projective relation between two types of nodes, space projection is performed to learn the \emph{projective representations} of nodes. 
Equation 1 formulates the projective representation learning process. In a bipartite network that contains projective information from homogeneous network $A$ to homogeneous network $B$, each node $A_i$ in network $A$ could learn a projective representation in network $B$, denoted as $Embd_{A_i\to B}$:
\begin{equation}Embd_{A_i\to B}= \frac{1}{N} \sum_{j=1}^N(Embd_{B_{j}}*w_{A_i B_j}) \end{equation}
Where $\to$ represents the projection relation in two spaces, \{${B_{N}}$\} is the complete set of objects in network $B$ that each $B_j$ in $B_N$ has $A_i\to B_j$. $Embd_{B_{j}}$ is the learned homogeneous representation of $B_j$, and $w_{A_i B_j}$ is the projective weight between nodes $A_i$ and $B_j$.

\paragraph{Final homogeneous network embedding}
Finally, the learned homogeneous network embeddings and  the bipartite network embeddings are concatenated to form the final homogeneous network embeddings in which  each node contains one homogeneous embedding and potentially several projective embeddings from bipartite subnetworks. The final heterogeneous embedding contains the information from different latent spaces, thus it can be regarded as an ensemble embedding that improves the robustness and generalization performance of a set of embeddings.

\section{Experiments}

\subsection{Dataset}
We construct an academic heterogeneous network, based on the dataset from AMiner Computer Science \cite{tang2008arnetminer}. The constructed network consists of two types of nodes: authors and papers, and three types of edges representing (i) authors coauthor with each other; (ii) authors write papers; (iii) papers cite other papers. After performing edge-based graph partition, two homogeneous subnetworks \textemdash the coauthor  network (Author-Author) and the citation network (Paper-Paper), and one bipartite network \textemdash writing network (Author-Paper), are generated.

\subsection{Baseline methods}
We compare our approach with several strong baseline methods including Line \citep{tang2015line}, DeepWalk \cite{perozzi2014deepwalk}, and Metapath2vec \cite{dong2017metapath}. 
The dimensions for LINE-based embeddings and the rest are 256 and 128 respectively. We set the size of negative samples to 5.
The number of random walks to start at each node in DeepWalk and Metapath2vec is 10, and the walk length is 40.

\subsection{Multi-label node classification}
We first evaluate the performance of GPSP on the multi-label classification task. We adopt the labeled dataset generated by the study~\cite{dong2017metapath}, which groups authors into 8 categories based on authors' research fields. Following the strategy in \cite{dong2017metapath}, we try to match this label set with the author embeddings, and get 103,024 successfully matched author embeddings with their labels. 

A SVM classifier is used to classify these  embeddings. To evaluate the robustness of our model, we compare  the performance of GPSP with competitors by varying the percentage of labeled data from 10\% to 90\%. 
The Micro-F1 and Macro-F1 scores are summarized in Table 1. 
GPSP-LINE and GSPS-DeepWalk substantially and consistently outperform the baseline methods by a noticeable margin on all experiments. 
Note that the metapath method~\cite{dong2017metapath} has a poor performance in the experiments, probably because that metapath2vec heavily relies on well structured paths that are difficult to obtain in many applications. 


\begin{table}[ht]

\begin{scriptsize}
\begin{center}
\begin{tabular}{  c| c| c c c c c} 
 \hline
 Metric & Model & 10\% & 30\% & 50\% & 70\%  & 90\%\\ 
 \hline
 \multirow{7}{5em}{Micro-F1} 
& LINE & 0.7062 & 0.7067& 0.7074& 0.7062& 0.7075 \\

& DeepWalk & 0.6992 & 0.7010& 0.6992& 0.6986& 0.6988 \\
 
& metapath2vec & 0.6546 & 0.6549& 0.6547& 0.6552& 0.6529 \\

& metapath2vec++ & 0.6692 & 0.6681& 0.6676& 0.6677& 0.6651 \\

& GPSP-LINE & \textbf{0.7512}&\textbf{0.7557}&\textbf{0.7564}&\textbf{0.7554}&\textbf{0.7552}\\
& GPSP-DeepWalk &\textbf{0.7275}&\textbf{0.7318}&\textbf{0.7324}&\textbf{0.7320}&\textbf{0.7318}\\
\hline
 
 \multirow{7}{5em}{Macro-F1}  
& LINE & 0.7032 & 0.7036& 0.7043& 0.7035&0.7036 \\

& DeepWalk & 0.6964 &0.6982&0.6965&0.6963&0.6961 \\

& metapath2vec & 0.6307 & 0.6313& 0.6322& 0.6328& 0.6301 \\

& metapath2vec++ & 0.6478 & 0.6473& 0.6478& 0.6473& 0.6445 \\
& GPSP-LINE &\textbf{0.7482}&\textbf{0.7527}&\textbf{0.7534}&\textbf{0.7526}&\textbf{0.7522}\\
& GPSP-DeepWalk &\textbf{0.7253}&\textbf{0.7290}&\textbf{0.7298}&\textbf{0.7295}&\textbf{0.7289}\\
 \hline
 
\end{tabular}
\end{center}
\end{scriptsize}
\caption{Multi-label node classification results}
\vspace{-0.3cm}
\end{table} 
\subsection{Node clustering}
To further evaluate the quality of the latent representations learned by GPSP, we also perform a node clustering task. We adopt simple K-means as our clustering algorithm, working on the learned latent representations. Here, $K$ is assigned to 8. The evaluation metric is normalized mutual information (NMI), which measures the mutual information  between the generated clusters and the labeled clusters.



The experimental results are demonstrated in Table 2. GPSP-DeepWalk achieves the best result, which improves 24\% in terms of NMI over the original DeepWalk method. 

\begin{table}[ht]
\begin{scriptsize}
\begin{center}
\begin{tabular}{cc c c c c} 
 \hline
 LINE &DeepWalk &metapath2v&metapath2v++& GPSP-LINE&GPSP-DeepWalk \\ 
 \hline
 0.2516  &0.2873&0.2403& 0.2470&\textbf{0.3118}&\textbf{0.3555}\\
\hline
\end{tabular}
\end{center}
\end{scriptsize}
\caption{Node clustering results (NMI scores)}
\vspace{-0.5cm}
\end{table}

\section{Conclusion}
A novel heterogeneous network embedding model, \textit{GPSP}, is proposed, which supports the representation learning of multiple types of nodes and edges. Extensive experiments show the superiority of GPSP by the benchmarks in two network mining tasks, node classification and clustering.
\bibliographystyle{ACM-Reference-Format}
\bibliography{bib} 

\end{document}